\documentclass[showpacs,preprintnumbers,amsmath,twocolumn,amssymb,prb]{revtex4}

\usepackage{amssymb}
\usepackage[dvips]{graphicx}
\usepackage{dcolumn}
\usepackage{bm}

\newcommand{\ud}{\,\mathrm{d}}
\newcommand{\ds}{\displaystyle}

\begin{document}

\title{Quantum dynamics of vortices in mesoscopic magnetic disks}

\author{R. Zarzuela$^{1}$, E. M. Chudnovsky$^{2}$, J. M. Hernandez$^{1}$, J. Tejada$^{1}$}
\affiliation{$^{1}$Departament de F\'{i}sica Fonamental, Facultat
de F\'{i}sica, Universitat de Barcelona, Avinguda Diagonal 645,
08028 Barcelona, Spain\\ $^{2}$Physics Department, Lehman College,
The City University of New York, 250 Bedford Park Boulevard West,
Bronx, NY 10468-1589, U.S.A.}

\date{\today}


\begin{abstract}
Model of quantum depinning of magnetic vortex cores from line defects in a disk geometry and under the application of an in-plane magnetic field 
has been developed within the framework of the Caldeira-Leggett theory. The corresponding instanton solutions are computed for several 
values of the magnetic field. Expressions for the crossover temperature $T_{c}$ and for the depinning rate $\Gamma(T)$ are obtained. 
Fitting of the theory parameters to experimental data is also presented.
\end{abstract}
\maketitle

\section{Introduction}

Macroscopic quantum tunneling of mesoscopic solid-state objects has been intensively studied in the past. Examples include single domain 
particles \cite{Chudnovsky1,Tejada,Vincent}, domain walls in magnets \cite{Egami,Chudnovsky2,Hong}, magnetic clusters \cite{Friedman,JM},
flux lines in type-II superconductors \cite{Blatter, Garcia-Santiago} and normal-superconducting interfaces in type-I superconductors 
\cite{Velez,Zarzuela1}. It is well known that micron-size circular disks made of soft ferromagnetic materials exhibit the \emph{vortex state} 
as the ground state of the system for a wide variety of diameters and thicknesses\cite{Cowburn,Shinjo,Hertel}. This essentially 
non-uniform magnetic configuration is characterized by the curling of the magnetization in the plane of the disk, leaving
virtually no magnetic ``charges''. The very weak uncompensated magnetic moment of the disk sticks out of a small area confined to
the vortex core (VC). The diameter of the core is comparable to the material exchange length\cite{Novosad1,Guslienko1} and, because 
of the strong exchange interaction among the out-of-plane spins in the VC, it behaves as an independent entity of mesoscopic size.

Recent experimental works have reported that the dynamics of the VC can be affected by the presence of structural defects in the sample
\cite{Shima,Compton,Zarzuela3,Burgess}. This is indicative of the elastic nature of the VC line, whose finite elasticity is provided by the exchange 
interaction\cite{Zarzuela2}. In Ref. \onlinecite{Zarzuela3} non-thermal magnetic relaxations under the application of an in-plane magnetic field 
are reported below $T=9$ K. It is attributed to the macroscopic quantum tunneling of the elastic VC line through pinning barriers when relaxing towards 
its equilibrium position. In such range of low temperatures only the softest dynamical mode can be activated, which corresponds to the gyrotropic 
motion of the vortex state. It consists of the spiral-like precessional motion of the VC as a whole\cite{Choe,Guslienko2,Guslienko3,Guslienko4,Lee} 
and it is intrinsically distinct from conventional spin wave excitations. It can also be viewed as the uniform precession of the magnetic moment 
of the disk due to the vortex. 

The aim of this paper is to study the mechanism of quantum tunneling of the elastic VC line through a pinning barrier during the gyrotropic motion. 
We focus our attention on line defects, which can be originated for instance by linear dislocations along the disk symmetry axis. This case may be relevant 
to experiments performed in Ref. \onlinecite{Zarzuela3} since linear defects provide the maximum pinning and, therefore, the VC line in the equilibrium state is 
likely to align locally with these defects. Such a situation would be similar to pinning of domain walls by interfaces and grain boundaries. Thus, 
we are considering the depinning of a small segment of the VC line from a line defect. The problem of quantum and thermal depinning of a massive elastic string 
trapped in a linear defect and subject to a small driving force was considered by Skvortsov \cite{Skvortsov}. The problem studied here is different as it 
involves gyrotropic motion of a massless vortex that is equivalent to the motion of a trapped charged string in a magnetic field \cite{Zarzuela2}. 
We study this problem with account of Caldeira-Leggett type dissipation.

The paper is structured as follows. In Sec. II the Lagrangian formalism of the generalized Thiele's equation is presented and 
Caldeira-Leggett theory is applied to obtain the depinning rate. The imaginary-time dynamical equation for instantons is derived in Section 
III and numerical solutions are computed. In Section IV the crossover temperature between the quantum 
and thermal regime is obtained. Discussion and fitting of the theory parameters (which is related to the pinning
potential) to experimental data are provided in Sec. V. Also final conclusions are included in this section.

\section{Elastic Thiele's lagrangian formalism and depinning rate}

\begin{figure}[htbp!]
\center
\includegraphics[scale=0.3]{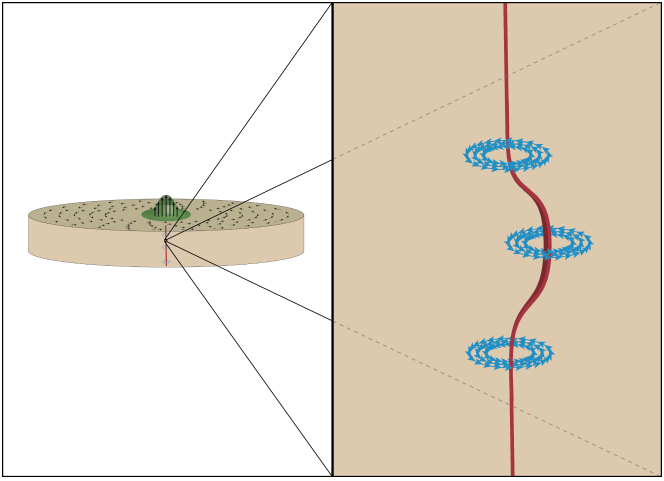}
\caption{(Color online) Vortex state and depinning via nucleation of the part of the VC line in a circular disk made of soft ferromagnetic material.}
\label{Fig1}
\end{figure}

In this paper we restrict ourselves to a circular disk geometry and to an applied in-plane magnetic field configuration. The VC line is pinned by 
the line defect going in the $Z$ direction (symmetry axis of the disk) at the center of the disk. The vortex line shall be described by the vector field $\vec{X}=(x,y)$, 
where $x(t,z)$ and $y(t,z)$ are coordinates of the center of the VC in the $XY$ plane. The dependence on the $Z$-coordinate emerges from the 
elastic nature of this magnetic structure. Figure \ref{Fig1} shows an sketch of the vortex line deformation due to pinning and its gyroscopic motion.

The softest dynamical mode of the VC, and hence of the whole vortex, originates from gyroscopic motion and it is described by the generalized 
Thiele's equation\cite{Zarzuela2}:
\begin{equation}
\label{Thiele}
\dot{\vec{X}}(t,z)\times\vec{\rho}_{G}
+\partial_{z}\vec{\Pi}_{z}+\nabla_{\vec{X}}\omega=0,
\end{equation}
where '\;$\dot{}$\;' means time derivative. 
The gyrovector density\cite{density} $\vec{\rho}_{G}=\rho_{G}\hat{e}_{z}$ is responsible for the gyroscopic motion 
of the VC and its modulus is given by $\rho_{G}=2\pi p n_{v} M_{s}/\gamma$, where $M_{s}$ is the saturation magnetization, $\gamma$ is the gyromagnetic 
ratio, $p=\pm1$ is the polarization of the VC and $n_{v}=\pm1$ is the vorticity of the magnetization of the disk. The potential energy density $\omega(\vec{X},\partial_{z}\vec{X})$ splits into the sum of 
two contributions, $\omega_{1}(\vec{X})$ and $\omega_{2}(\partial_{z}\vec{X})$. The latter is the elastic energy term, 
$\ds \omega_{2}(\partial_{z}\vec{X})=\frac{1}{2}\lambda\left(\frac{\partial\vec{X}}{\partial z}\right)^{2}$, which is provided by the exchange 
interaction. The elastic constant is given by $\lambda=2\pi A\ln(R/\Delta_{0})$, where $R$ is the radius of the disk, $A$ is the exchange constant and
$\Delta_{0}=\sqrt{A/M_{s}^2}$ is the exchange length of the ferromagnetic material. Finally, $\vec{\Pi}_{z}=-\delta\omega/\delta
(\partial_{z}\vec{X})=-\lambda\partial_{z}\vec{X}$ is the generalized momentum density with respect to $Z$. Consequently, the generalized 
Thiele's equation becomes
\begin{equation}
\label{Thiele2}
\dot{\vec{X}}(t,z)\times\vec{\rho}_{G}-\lambda\partial^{2}_{z}\vec{X}(t,z)+\nabla_{\vec{X}}\omega=0
\end{equation}

Let $L$ be the thickness of the circular disk. The Lagrangian corresponding to the above equation is given by\cite{Zarzuela2}
\begin{align}
\label{Lagrangian}
 \mathcal{L}[t,\vec{X},\dot{\vec{X}},\partial_{z}\vec{X}]
 =\int_{0}^{L}\ud z&\;\Bigg\{\dot{\vec{X}}\cdot\vec{A}_{\rho_{G}}
-\omega(\vec{X},\partial_{z}\vec{X})\Bigg\},
\end{align}
where $\ds \vec{A}_{\rho_{G}}=\rho_{G}y\hat{e}_{x}$ is the gyrovector potential in a convenient gauge\cite{gauge}. The depinning rate 
at a temperature $T$, $\Gamma(T)=A(T)\exp{[-B(T)]}$, is obtained by performing the imaginary-time path integral\cite{MQTMM}
\begin{equation}
 \label{Dep_rate}
\int D\{x\}\int D\{y\}\exp{\Bigg[\frac{-1}{\hbar}\oint \ud\tau\mathcal{L}_{E}\Bigg]}
\end{equation}
over $\vec{X}_{\tau}\equiv\vec{X}(\tau,z)$ trajectories, which are periodic in $\tau$ with period $\hbar/k_{B}T$. Notice that $\tau=it$ is the imaginary time 
and $\mathcal{L}_{E}$ is the Euclidean version of Eq. \eqref{Lagrangian}. That is,
\begin{align}
\label{ELagrangian}
 \mathcal{L}_{E}[\tau,\vec{X}_{\tau},\dot{\vec{X}}_{\tau},\partial_{z}\vec{X}_{\tau}]
 =\int_{0}^{L}\ud z&\;\Bigg\{-i\dot{\vec{X}}_{\tau}\cdot\vec{A}_{\rho_{G}}\nonumber\\
&+\omega(\vec{X}_{\tau},\partial_{z}\vec{X}_{\tau})\Bigg\},
\end{align}

The energy density $\omega_{1}(\vec{X}_{\tau})$ splits into the sum of three terms: The first one, $\omega_{XY}(\vec{X}_{\tau})$, represents the 
sum of the magnetostatic and exchange contributions in the $z$-cross-section, whose dependence on the vortex core coordinates is 
$\omega_{XY}(\vec{X}_{\tau})\sim\vec{X}_{\tau}^ 2$ for small displacements\cite{Zarzuela2}. The second term, $\omega_{dep}(\vec{X}_{\tau})$, 
represents the pinning energy density associated to the line defect. Recent experimental works 
have reported an even quartic dependence of pinning potentials on the VC coordinates for small displacements in permalloy rings\cite{Bedau}. 
Consequently, it is legitimate to take the following functional dependence for the sum of both terms:
\begin{equation}
 \label{potential}
 (\omega_{XY}+\omega_{dep})(\vec{X}_{\tau})=\frac{1}{2}\kappa\left(x_{\tau}^{2}+y_{\tau}^{2}\right)-\frac{1}{4}\beta p_{4}(x_{\tau},y_{\tau})
\end{equation}
where $(\kappa,\beta)$ are the parameters of our model and $p_{4}(x,y)$ is a linear combination of monomials of degree four on variables $x$ and $y$. 
The last term is the Zeeman energy density, which is given by\cite{Zarzuela2} $\omega_{Z}(\vec{X}_{\tau})=-\mu\left[\hat{z}\times\vec{H}_{in}\right]\cdot\vec{X}_{\tau}$ 
-with $\mu=(2\pi/3)M_{s} n_v R$- for small displacements. The latter correspond to the application of a weak in-plane magnetic field $\vec{H}_{in}$. In what follows, 
$\vec{H}_{in}=-H\hat{e}_{y}$ is applied along the $Y$ direction.

The simple dependence $p_{4}(x_{\tau},y_{\tau})= x_{\tau}^{4}$ keeps the main features of the pinning potential (see Section V). We also neglect the elastic term 
$\frac{1}{2}\lambda\left(\frac{\partial y_{\tau}}{\partial z}\right)^{2}$. From all these considerations, the Lagrangian \eqref{ELagrangian} becomes

\begin{align}
\label{ELagrangian2}
 \mathcal{L}_{E}[\tau,\vec{X}_{\tau},&\dot{\vec{X}}_{\tau},\partial_{z}\vec{X}_{\tau}]
 =\int_{0}^{L}\ud z\;\Bigg\{-i\rho_{G}y_{\tau}\dot{x}_{\tau}-\mu h x_{\tau}\nonumber\\
&+\frac{\kappa}{2} x_{\tau}^{2}+
\frac{\kappa}{2} y_{\tau}^{2}-\frac{\beta}{4}x_{\tau}^{4}+\frac{\lambda}{2}\left(\frac{\partial x_{\tau}}{\partial z}\right)^{2}\Bigg\}
\end{align}

Finally, Gaussian integration over $y_{\tau}$ reduces Eq. \eqref{Dep_rate} to
\begin{equation}
 \label{Dep_rate2}
\int D\{x\}\exp{\Bigg[\frac{-1}{\hbar}\oint \ud\tau\mathcal{L}_{E,eff}\Bigg]}
\end{equation}
with
\begin{align}
\label{ELagrangian_eff}
 \mathcal{L}_{E,eff}[\tau,x_{\tau},\dot{x}_{\tau},&\partial_{z}x_{\tau}]
 =\int_{0}^{L}\ud z\;\Bigg\{\frac{1}{2}\left(\frac{\rho_{G}^{2}}{\kappa}\right)\dot{x}^{2}_{\tau}-\mu h x_{\tau}\nonumber\\
&+\frac{\kappa}{2} x_{\tau}^{2}-\frac{\beta}{4}x_{\tau}^{4}+\frac{\lambda}{2}\left(\frac{\partial x_{\tau}}{\partial z}\right)^{2}\Bigg\}
\end{align}

Within the framework of the Caldeira-Leggett theory\cite{Caldeira}, dissipation is taken into account by adding a term
\begin{equation}
 \frac{\eta}{4\pi}\int_{0}^{L}\ud z\oint\ud\tau\int_{\mathbb{R}}\ud\tau_{1}\frac{(x_{\tau}(\tau,z)-x_{\tau}(\tau_{1},z))^{2}}{(\tau-\tau_{1})^{2}}
\end{equation}
to the action of Eq. \eqref{Dep_rate2}. The dissipative constant $\eta$ is related to the damping of the magnetic vortex core\cite{MQTMM} and 
Ref. \onlinecite{Guslienko2} shows that $\eta\simeq3\alpha_{LLG}|\rho_{G}|$, with $\alpha_{LLG}$ being the Gilbert damping parameter. Introducing 
dimensionless variables $\bar{\tau}=(\kappa/\sqrt{2}|\rho_{G}|)\tau$, $\bar{z}=(\kappa/2\lambda)^{1/2}z$ and $u=(2\beta/\kappa)^{1/2}x_{\tau}$, 
the depinning exponent becomes
\begin{multline}
\label{Dep_exp}
B(T,h)=\frac{|\rho_{G}|\sqrt{\lambda\kappa}}{2\hbar \beta}\int\ud\bar{z}\oint\ud\bar{\tau}\Bigg[\frac{1}{2}\dot{u}^2+\frac{1}{2}(u')^2+V(u,h)\\
+\frac{\eta}{2\sqrt{2}\pi|\rho_{G}|}\int_{\mathbb{R}}\ud \bar{\tau}_{1}\frac{(u(\bar{\tau},\bar{z})-u(\bar{\tau}_{1},\bar{z}))^2}{(\bar{\tau}-\bar{\tau}_{1})^2}\Bigg]
\end{multline}
where '\;$'$\;' means derivative with respect to $\bar{z}$, $\ds V(u,h)=-h u+u^2-\frac{u^4}{4}$ is the normalized energy potential 
and $h=2\sqrt{2\beta/\kappa^3}\mu H$. Let $u_{0}(h)$ be the relative minimum of $V$ for a fixed value of $h$. We reescale 
the energy potential $V(u,h)\rightarrow V(u,h):=V(u_{0}(h)+u,h)-V(u_{0}(h),h)$ and the variable 
$u\rightarrow u_{0}(h)+u$ so that we obtain $\ds V(u,h)=u^2\left(\left(1-\frac{3}{2}u_{0}^{2}(h)\right)-u_{0}(h)u
-\frac{1}{4}u^2\right)$. 

\section{Instantons of the dissipative 1+1 model}

Quantum depinning of the VC line is given by the instanton solution of the Euler-Lagrange equations of motion of the 1+1 field theory described 
by Eq. \eqref{Dep_exp}. This gives
\begin{multline}
 \label{EL_eq}
\ddot{u}+u''-\big(2-3u_{0}^2(h)\big)u+3u_{0}(h)u^2+u^3-\\
\frac{\sqrt{2}}{\pi}\frac{\eta}{|\rho_{G}|}\int_{\mathbb{R}}\ud\bar{\tau}_{1}
\frac{u(\bar{\tau},\bar{z})-u(\bar{\tau}_{1},\bar{z})}{(\bar{\tau}-\bar{\tau}_{1})^2}=0
\end{multline}
with boundary conditions 
\begin{align}
 u(-\Omega/2,\bar{z})=u(\Omega/2,\bar{z}) & &\bar{z}\in\mathbb{R}\nonumber\\
 \max_{\bar{\tau}\in[-\Omega/2,\Omega/2]}u(\bar{\tau},\bar{z})=u(0,\bar{z})& &\bar{z}\in\mathbb{R}
\end{align}
that must be periodic on the imaginary time $\bar{\tau}$ with the period $\ds \Omega=\frac{\kappa}{\sqrt{2}|\rho_{G}|}\frac{\hbar}{k_{B}T}$. 
This equation cannot be solved analytically, so we must proceed by means of numerical methods. Notice that in the computation of instantons we 
can safely extend the integration over $\bar{z}$ in Eq. \eqref{Dep_exp} on the the whole set of real numbers.
\subsection{Zero temperature}
In this case we apply the 2D Fourier transform 
\begin{equation}
 \hat{u}(\omega,\theta)=\frac{1}{2\pi}\int_{\mathbb{R}^2}\ud\bar{\tau}\ud\bar{z}\;u(\bar{\tau},\bar{z})e^{i(\omega\bar{\tau}+\theta\bar{z})}
\end{equation}
to Eq. \eqref{EL_eq} and obtain
\begin{multline}
 \label{FT0eq}
 \hat{u}(\omega,\theta)=\frac{1}{\omega^2+\theta^2+\sqrt{2}|\omega|\eta/|\rho_{G}|+2-3u_0^2(h)}\Bigg(\frac{3u_0(h)}{2\pi}\\
\times\int_{\mathbb{R}^2}\!\!\!\ud\omega_{1}\ud\theta_{1}\hat{u}(\omega_1,\theta_1)\hat{u}(\omega-\omega_1,\theta-\theta_1)+\frac{1}{(2\pi)^2}\times\\
\int_{\mathbb{R}^4}\!\!\!\ud^2\vec{\omega}\ud^2\vec{\theta}\;\hat{u}(\omega_2,\theta_2)\hat{u}(\omega_1-\omega_2,\theta_1-\theta_2)
\hat{u}(\omega-\omega_1,\theta-\theta_1)\Bigg)
\end{multline}
which is an integral equation for $\hat{u}$. The depinning exponent \eqref{Dep_exp} in the Fourier space becomes 
\begin{multline}
 \label{BT0}
B(T=0,h)=\frac{|\rho_{G}|\sqrt{\kappa\lambda}}{2\hbar\beta}\Bigg[\int_{\mathbb{R}^2}\ud\omega\ud\theta\;\hat{u}(\omega,\theta)\hat{u}(-\omega,
-\theta)\\\Bigg(\left(1-\frac{3}{2}u_{0}^{2}(h)\right)+\frac{\omega^2+\theta^2}{2}+\frac{|\omega|}{\sqrt{2}}\frac{\eta}{|\rho_{G}|}\Bigg)-\frac{u_{0}(h)}{2\pi}
\int_{\mathbb{R}^4}\ud^2\vec{\omega}\ud^2\vec{\theta}\\
\hat{u}(\omega_1,\theta_1)\hat{u}(\omega_2,\theta_2)\hat{u}(-\omega_1-\omega_2,
-\theta_1-\theta_2)-\frac{1}{(4\pi)^2}\int_{\mathbb{R}^6}\ud^3\vec{\omega}\ud^3\vec{\theta}\\
\hat{u}(\omega_1,\theta_1)\hat{u}(\omega_2,\theta_2)
\hat{u}(\omega_3,\theta_3)\hat{u}(-\omega_1-\omega_2-\omega_3,-\theta_1-\theta_2-\theta_3)\Bigg].
\end{multline}

The zero-temperature instanton is computed using an algorithm that is a field-theory extension of
the algorithm introduced in Refs. \onlinecite{Chang}, \onlinecite{Waxman} for the problem of dissipative quantum tunneling of a particle: To begin 
with, we introduce the operator
\begin{multline}
 \label{OT0}
O(\lambda,\alpha,\hat{u}(\omega,\theta),h)=
\frac{1}{\omega^2+\theta^2+\sqrt{2}|\omega|\eta/|\rho_{G}|+2-3u_0^2(h)}\times\\
\Bigg(\lambda\int_{\mathbb{R}^2}\!\!\ud\omega_{1}\ud\theta_{1}\hat{u}(\omega_1,\theta_1)\hat{u}(\omega-\omega_1,\theta-\theta_1)+\\
\alpha\int_{\mathbb{R}^4}\!\!\ud^2\vec{\omega}\ud^2\vec{\theta}\;\hat{u}(\omega_2,\theta_2)\hat{u}(\omega_1-\omega_2,\theta_1-\theta_2)
\hat{u}(\omega-\omega_1,\theta-\theta_1)\Bigg).
\end{multline}

Secondly, it is important to point out the \emph{scaling property} of this operator because it will be used in the computation of Eq. \eqref{BT0}: 
Given any triplet $(\lambda_{0},\alpha_{0},\hat{u}_{0}(\omega,\theta))$ satisfying the identity \eqref{FT0eq}, so will any other triplet 
$(\lambda_{1},\alpha_{1},\hat{u}_{1}(\omega,\theta))$ provided that
 \begin{align}
\label{coef_rel}
\hat{u}_1(\omega,\theta)&=\chi \hat{u}_0(\omega,\theta)\\
\lambda_1&=\lambda_0/\chi\\
\alpha_1&=\alpha_0/\chi^2,
\end{align}
where $\chi$ is a constant. This means that if we are able to find a solution $(\lambda_{1},\alpha_{1},\hat{u}_{1}(\omega,\theta))$ for 
arbitrary parameters $(\lambda_1,\alpha_1)$, then we can obtain the solution corresponding to the pair $(\lambda_0,\alpha_0)$ simply by 
rescaling $\hat{u}_1(\omega,\theta)$ by a factor $\chi=\lambda_1/\lambda_0$ as long as $(\lambda_1/\lambda_0)^2=\alpha_1/\alpha_0$ is verified. 

The algorithm consists of the following steps:
\begin{enumerate}
 \item Start with an initial $(\lambda_0,\alpha_0,\hat{u}_0(\omega,\theta))$.
 \item Let $\hat{u}_{1}(\omega,\theta)=O(\lambda_0,\alpha_0,\hat{u}_0(\omega,\theta),h)$.
 \item Calculate $\lambda_1=\lambda_0/\chi^2, \alpha_1=\alpha_0/\chi^3$, where $\chi=\hat{u}_{1}(\vec{0})/\hat{u}_{0}(\vec{0})$.
 \item Find $\hat{u}_{2}(\omega,\theta)=O(\lambda_1,\alpha_1,\hat{u}_{1}(\omega,\theta),h)$.
 \item Repeat steps (2)-(4) until the successive difference satisfies a preset convergence criterion.
\end{enumerate}
The output is the triplet $(\lambda_n,\alpha_n,\hat{u}_n(\omega,\theta))$. The final step consists of reescaling $\hat{u}_{n}$ to obtain the 
solution corresponding to the pair $(\lambda,\alpha)=(3u_{0}(h)/2\pi,1/(2\pi)^{2})$: from the scaling property we know that the reescaling 
rules of the $\lambda$- and $\alpha$- terms of Eq. \eqref{FT0eq} are different. Thus, to obtain an accurate approximation of the instanton solution 
we have split $\hat{u}(\omega,\theta)$ into the sum of two functions $\hat{u}_1(\omega,\theta)$ and $\hat{u}_2(\omega,\theta)$ 
in the above algorithm, and calculated their next iteration by means of the $\lambda$-term, respectively the $\alpha$-term of the operator \eqref{OT0}. Finally, we 
rescale $\hat{u}_{1}$ by a factor $2\pi\lambda_{n}/3u_{0}(h)$ and $\hat{u}_2$ by a factor $2\pi\sqrt{\alpha_{n}}$. The depinning rate is 
calculated evaluating Eq. \eqref{BT0} at this solution.

\subsection{Non-zero temperature}
In the $T\neq0$ case, taking into account the finite periodicity on $\bar{\tau}$ we consider a solution of the type
\begin{equation}
 u(\bar{\tau},\bar{z})=\sum_{n\in\mathbb{Z}} u_{n}(\bar{z})e^{-i\omega_{n}\bar{\tau}}
\end{equation}
with $\ds\omega_{n}=\frac{2\pi n}{\Omega}$ for all $n\in\mathbb{Z}$. Introducing this functional dependence into Eq. \eqref{EL_eq} and applying a 1D 
Fourier transform we obtain
\begin{multline}
 \label{FTneq0}
\hat{u}_{n}(\theta)=\frac{1}{\omega_{n}^2+\theta^2+\sqrt{2}|\omega_{n}|\eta/|\rho_{G}|+2-3u_0^2(h)}\times\\
\Bigg(\frac{3u_0(h)}{\sqrt{2\pi}}\sum_{p\in\mathbb{Z}}\int_{\mathbb{R}}\ud\theta_1\hat{u}_{p}(\theta_1)\hat{u}_{n-p}(\theta-\theta_1)+\\
\frac{1}{2\pi}\sum_{p,q\in\mathbb{Z}}\int_{\mathbb{R}^2}\!\!\ud^2\vec{\theta}\;\hat{u}_{p}(\theta_2)\hat{u}_{q}(\theta_1-\theta_2)
\hat{u}_{n-p-q}(\theta-\theta_1)\Bigg)
\end{multline}
which is an integral equation for the set $\{\hat{u}_{n}\}_{n\in\mathbb{Z}}$ of Fourier coefficients. The depinning exponent \eqref{Dep_exp} 
in the Fourier space becomes
\begin{multline}
 \label{BTneq0}
B(T>0,h)=\frac{|\rho_{G}|\sqrt{\kappa\lambda}}{2\hbar\beta}\Bigg[\sum_{n\in\mathbb{Z}}\int_{\mathbb{R}}\ud\theta\;\hat{u}_{n}(\theta)
\hat{u}_{-n}(-\theta)\\
\Bigg(\left(1-\frac{3}{2}u_{0}^{2}(h)\right)+\frac{\omega_{n}^2+\theta^2}{2}+\frac{|\omega_{n}|}{\sqrt{2}}\frac{\eta}{|\rho_{G}|}\Bigg)-\frac{u_{0}(h)}{\sqrt{2\pi}}
\sum_{n,m\in\mathbb{Z}}\int_{\mathbb{R}^2}\ud^2\vec{\theta}\\
\hat{u}_{n}(\theta_1)\hat{u}_{m}(\theta_2)\hat{u}_{-n-m}(-\theta_1-\theta_2)-\frac{1}{8\pi}\sum_{n,m,l\in\mathbb{Z}}\int_{\mathbb{R}^3}\ud^3\vec{\theta}\\
\hat{u}_{n}(\theta_1)\hat{u}_{m}(\theta_2)\hat{u}_{l}(\theta_3)\hat{u}_{-n-m-l}(-\theta_1-\theta_2-\theta_3)\Bigg]\Omega
\end{multline}
The numerical algorithm is analogous to the one used in the zero-temperature case, but taking into account the reescaling of 
$\{\hat{u}_{p}^{1}\}_{p\in\mathbb{Z}}$ by a factor $\sqrt{2\pi}\lambda_{n}/3u_{0}(h)$ and $\{\hat{u}_{p}^{2}\}_{p\in\mathbb{Z}}$ 
by a factor $\sqrt{2\pi\alpha_{n}}$ in the last step of the calculations.

\begin{figure}[htbp!]
\center
\includegraphics[scale=0.29]{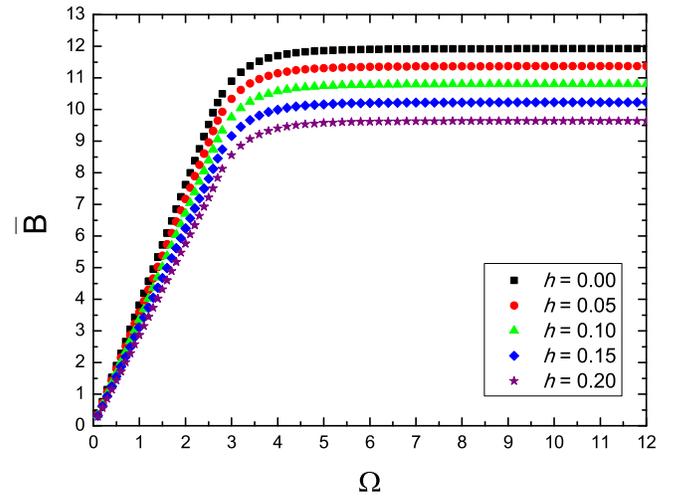}
\caption{(Color online) Temperature dependence of the depinning rate: normalized action $\ds \frac{2\hbar\beta}{|\rho_{G}|\sqrt{\kappa\lambda}}B(T)$ 
versus $\Omega$ at different values of the parameter $h$.}
\label{Fig2}
\end{figure}

Fig. \ref{Fig2} shows the normalized action $\ds \bar{B}(T)=\frac{2\hbar\beta}{|\rho_{G}|\sqrt{\kappa\lambda}}B(T)$ as a function of $\Omega$ at different 
values of the parameter $h$. In the simulations we have taken the standard value $\alpha_{LLG}=0.008$ for bulk Permalloy\cite{Guslienko2}.

\section{Crossover temperature}

The crossover temperature determines the transition from thermal to quantum tunneling relaxation regimes. It can be computed by means of theory of 
phase transitions\cite{Larkin}: above $T_{c}$, the instanton solution minimizing Eq. \eqref{Dep_exp} is a $\bar{\tau}$-independent function 
$u(\bar{\tau},\bar{z},h)=\bar{u}_{0}(\bar{z},h)$, whereas just below $T_{c}$ the instanton solution can be split into the sum 
of $\bar{u}_{0}$ and a small perturbation depending on $\bar{\tau}$,
\begin{equation}
\label{CTdep}
 u(\bar{\tau},\bar{z},h)=\bar{u}_{0}(\bar{z},h)+\bar{u}_{1}(\bar{z},h)\cos\left(\frac{2\pi}{\Omega}\bar{\tau}\right)
\end{equation}
The depinning exponent \eqref{Dep_exp} is proportional to
\begin{equation}
 \int_{\mathbb{R}}\ud\bar{z}\;\Phi(\bar{z};\bar{u}_{1},\bar{u}_{1}')
\end{equation}
where $\Phi$ is the spatial action density. Introducing the expansion \eqref{CTdep} into Eq. \eqref{Dep_exp} we obtain the following expansion
\begin{align}
 \Phi(\bar{z};\bar{u}_{1},&\bar{u}_{1}')=\left[\frac{1}{2}(\bar{u}_{0}')^{2}+V(\bar{u}_{0},h)\right]\Omega\;+\nonumber\\
&\frac{\Omega}{4}(\bar{u}_{1}')^2+\Lambda \bar{u}_{1}^{2}+O(4)
\end{align}
with 
\begin{equation}
\Lambda=\frac{\Omega}{4}V''(\bar{u}_{0},h)+\frac{\pi^2}{\Omega}+\frac{\pi}{\sqrt{2}}\frac{\eta}{|\rho_{G}|}
\end{equation}
If $\Lambda>0$ the only pair $(\bar{u}_{1},\bar{u}_{1}')$ minimizing $\Phi$ is $\bar{u}_{1}\equiv0$. The crossover temperature is then defined by 
the equation $\min_{\bar{z}\in\mathbb{R}}\Lambda=0$, that is
\begin{equation}
\label{Omega_c}
\frac{\Omega_{c}}{4}\min_{\bar{z}\in\mathbb{R}}V''(\bar{u}_{0},h)+\frac{\pi^2}{\Omega_{c}}+\frac{\pi}{\sqrt{2}}\frac{\eta}{|\rho_{G}|}=0
\end{equation}
The equation of motion for a $\bar{\tau}$-independent instanton is
\begin{equation}
\label{indepins}
\bar{u}_{0}''-(2-3u_{0}^{2}(h))\bar{u}_{0}+3u_{0}(h)\bar{u}_{0}^{2}+\bar{u}_{0}^{3}=0
\end{equation}
with boundary conditions: $\bar{u}_{0}\rightarrow0$ at $|\bar{z}|\rightarrow\infty$ and $\bar{u}_{0}(0,h)=-2u_{0}(h)+
\sqrt{4-2u_{0}^{2}(h)}\equiv w(h)$, which is the width of the potential. Consequently, 
\begin{align}
 \min_{\bar{z}\in\mathbb{R}}&V''(\bar{u}_{0}(\bar{z},h),h)=\min_{\bar{u}_{0}\in[0,w(h)]}V''(\bar{u}_{0},h)=\nonumber\\
&\min_{\bar{u}_{0}\in[0,w(h)]}\Big\{(2-3u_{0}^{2}(h))-6u_{0}(h)\bar{u}_{0}-3\bar{u}_{0}^{2}\Big\}=\nonumber\\
&-10+3u_{0}^{2}(h)+6u_{0}(h)\sqrt{f(h)}
\end{align}
with $f(h)=4-2u_{0}^{2}(h)$. Solving the quadratic equation for $T_{c}$ given by Eq. \eqref{Omega_c} we obtain the crossover temperature
\begin{equation}
\label{Tc}
T_{c}(h)\!=\!\frac{\hbar\kappa}{4\pi k_{B}|\rho_{G}|}\!\Bigg[\!\sqrt{8\!+\!3f(h)\!-\!12u_0(h)\!\sqrt{f(h)}\!+\!\frac{\eta^2}
{\rho_{G}^2}}\!-\!\frac{\eta}{|\rho_{G}|}\!\Bigg]
\end{equation}

\begin{figure}[htbp!]
\center
\includegraphics[scale=0.29]{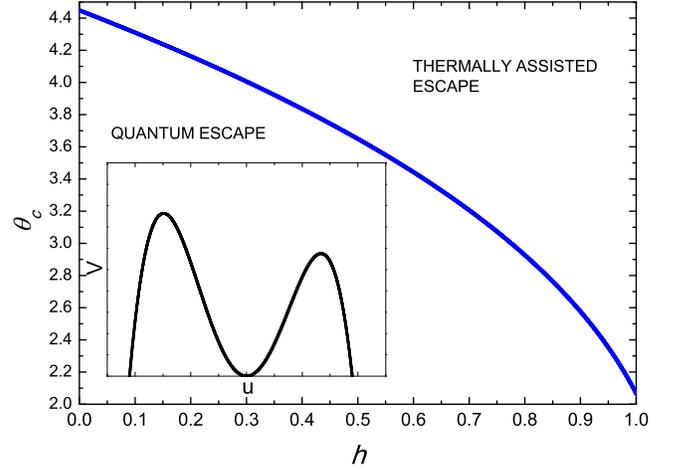}
\caption{(Color online) Field dependence of the dimensionless crossover temperature $\ds \theta_{c}=\frac{4\pi k_{B}|\rho_{G}|}{\hbar\kappa}T_{c}$. 
(Inset) Sketch of the potential $V(u,h=0.1)$.}
\label{Fig3}
\end{figure}

Figure \ref{Fig3} shows the dependence of the dimensionless crossover temperature $\ds \theta_{c}=\frac{4\pi k_{B}|\rho_{G}|}{\hbar\kappa}
T_{c}$ on the generalized magnetic field $h$.

\section{Discussion and parameters fitting}

For a given value of the generalized field $h$, in Fig. \ref{Fig2} we clearly distinguish two regimes in the dependence of the normalized action on 
$\Omega$: above $\Omega_{c}(h)$ the normalized action tends to a constant value, whereas below it the normalized action is linear with $\Omega$. 
Notice that the transition from the linear to the constant regime is smooth (that is, of second-order type). Above $T_{c}$ the depinning rate becomes
\begin{multline}
B(T>T_{c},h)=\frac{|\rho_{G}|\sqrt{\lambda\kappa}}{2\hbar \beta}\int\ud\bar{z}\left[\frac{1}{2}(\bar{u}_{0}')^2+V(\bar{u}_{0},h)\right]\Omega
\end{multline}
with $\bar{u}_{0}$ being the $\bar{\tau}$-independent instanton. By means of Eq. \eqref{indepins} this expression can be rewritten as\cite{MQTMM}
\begin{multline}
B(T>T_{c},h)=\frac{|\rho_{G}|\sqrt{\lambda\kappa}}{2\hbar \beta}\left[2\sqrt{2}\int_{0}^{w(h)}\ud \bar{u}_{0}\sqrt{V(\bar{u}_{0},h)}\right]\Omega
\end{multline}
and, consequently, the slope of the normalized action $\bar{B}(\Omega)$ is equal to $2\sqrt{2}\int_{0}^{w(h)}\ud \bar{u}_{0}\sqrt{V(\bar{u}_{0},h)}$, 
which can be evaluated analytically. At all values of the generalized field $h$, the numerical slope calculated from Fig. \ref{Fig2} coincides with the analytical one within the numerical error of our simulations. This is 
indicative of the robustness of our algorithm.

Quantum effects reported in Ref. \onlinecite{Zarzuela3} can be understood as being plausibly due to the depinning from line defects present in the disk. 
The size of the defects needs to exceed the nucleation length in order to pin the VC, but not to be as long as the thickness of the disk.  Pinning of extended parts of the VC line 
by line defects would be justified by the fact that linear defects provide the strongest pinning so that the VC line, or at least some segments of it, would naturally 
fall into such traps. Consequently, we can test out our model on the experimental results obtained in Ref. \onlinecite{Zarzuela3}. The crossover temperature is relevant to the roughness of the fine-scale potential landscape due to linear defects at the bottom of the potential 
well created by the external and dipolar fields. Above $T_c$ vortices diffuse in this potential by thermal activation, whereas below $T_c$ they 
diffuse by quantum tunneling. This must determine the temperature dependence (independence) of the magnetic viscosity. $T_c$ is, therefore, the
measure of the fine-scale barriers due to linear defects. It can be measured experimentally and help to extract the width of the pinning potential.

\begin{figure}[htbp!]
\center
\includegraphics[scale=0.29]{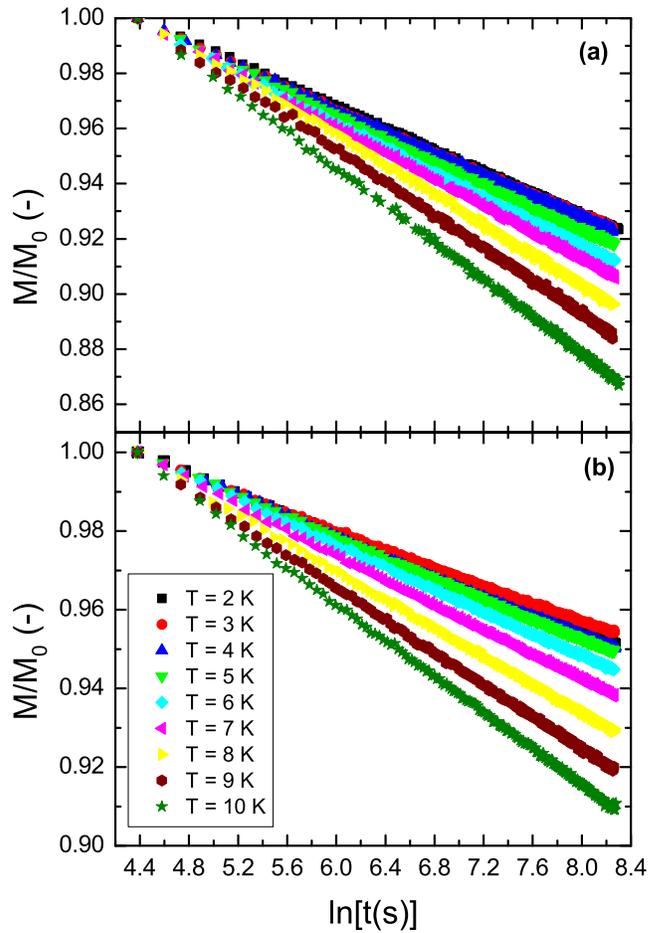}
\caption{(Color online) Relaxation measurements of magnetic vortices from the remnant state to equilibrium for samples $(L,R)=(95,750)$ nm 
(subfig (a)) and $(L,R)=(60,750)$ nm (subfig (b)).
}
\label{Fig4}
\end{figure}

Now we proceed to obtain estimates of the model parameters $(\kappa,\beta)$ by fitting our model to experimental data: Figure \ref{Fig4} 
shows new magnetic relaxation measurements of permalloy disks in the vortex state from the remnant state to equilibrium (zero magnetization). 
The radius of these disks is $R=0.75\textrm{ }\mu$m and their thickness is $L=95$ nm (subfig. \ref{Fig4}a) and $L=60$ nm (subfig. \ref{Fig4}b). 
A concise description of the experimental set-up and sample preparation can be found in Ref. \onlinecite{Zarzuela3}. Notice that for both 
samples the magnetization depends logarithmically on time during the relaxation process.

\begin{figure}[htbp!]
\center
\includegraphics[scale=0.29]{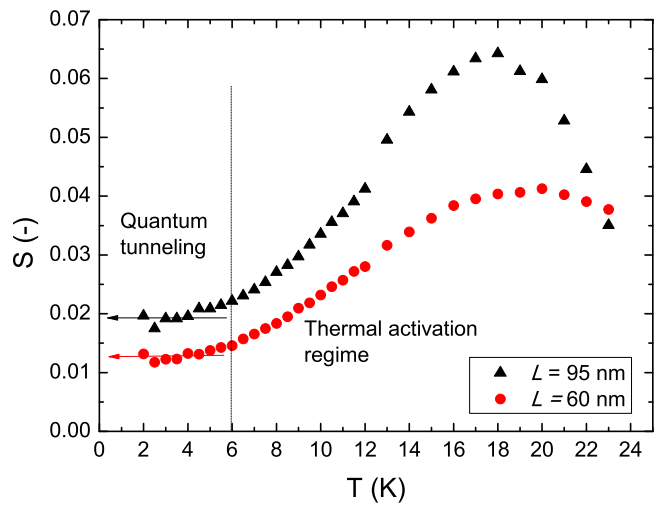}
\caption{(Color online) Magnetic viscosity versus temperature for both samples. The arrows point towards the non-zero value of the plateaus.}
\label{Fig5}
\end{figure}

Magnetic viscosity of these relaxation measurements is computed by means of the formula\cite{MQTMM}:
\begin{equation}
S(T)=-\frac{1}{M_{0}}\frac{\partial M}{\partial\ln t},
\end{equation}
where $M_{0}$ is the initial magnetization point. That is, the viscosity at zero field is obtained computing the slopes of the normalized 
magnetization curves. Fig \ref{Fig5} shows the magnetic viscosity as a function of temperature for both samples. Below $T_{c}=6$ K, magnetic 
viscosity reaches a plateau with non-zero value. Above $T_{c}$, magnetic viscosity increases up to a certain temperature, from which it decreases 
again. The existence of the plateau is the evidence of underbarrier quantum tunneling phenomena. The increase of viscosity with temperature above the
crossover temperature is due to thermal activation over the pinning barriers. Finally, the drop of the magnetic viscosity is in agreement with the 
loss of magnetic irreversibility in our systems\cite{Zarzuela3}. On the other hand, the fact that the crossover temperature $T_{c}$ is independent of 
the thickness of the disks upholds our hypothesis that just a small portion of the VC line takes part in the tunneling process via an elastic deformation.

Notice that the depinning rate should not exceed $30-40$ in order for the tunneling to occur on a reasonable time scale. The estimates of the 
parameters $(\kappa,\beta)$ are obtained fitting Eq. \eqref{Tc} and Eq. \eqref{BT0} to the values $T_{c}\sim6$ K, respectively $B(T=0,h=0)\sim30$ at zero field. 
Considering the experimental values $A=1.3\cdot10^{-11}$ J/m and $M_{s}=7.5\cdot10^{5}$ A/m for permalloy, we obtain
\begin{eqnarray}
 \kappa\sim5.9\cdot10^{7}\textrm{ J/m}^{3},& & \beta\sim6.9\cdot10^{27}\textrm{ J/m}^{5}
\end{eqnarray}
from which we can determine the width of the quartic potential, $\ds w=\sqrt{2\kappa/\beta}\sim0.13$ nm. This value is compatible with the width of the potential 
provided by a linear dislocation

In conclusion, we have studied quantum escape from a line defect of the VC line in a disk made of a soft ferromagnetic material. In the case of 
permalloy disks, experimental results let us conclude that the depinning process occurs in steps about 0.13 nm, which corresponds to the width of the 
energy potential.

\section{Acknowledgments}

The authors acknowledge V. Novosad for providing the samples discussed in the paper and Grup de Din\`{a}mica Financera de la UB for the 
use of their computing facilities. The work of R.Z. has been financially supported by the Ministerio de Ciencia e Innovaci\'{o}n de Espa\~na. 
J.T. acknowledges financial support from ICREA Academia. The work at the University of Barcelona was supported by the Spanish Government 
Project No. MAT2011-23698. The work of E.M.C. at Lehman College has been supported by the U.S. National Science Foundation through grant 
No. DMR-1161571.
\bibliography{Theo_vortices}

\begin{thebibliography}{36}
\bibitem[{gau()}]{gauge}
\bibinfo{note}{In Ref. \onlinecite{Zarzuela2} the symmetric gauge is used
  instead of this one. In both cases the gyrovector potential verifies the
  identity $\nabla_{\vec{X}}\times\vec{A}_{\rho_{G}}=-\vec{\rho}_{G}$.}
\end{thebibliography}

\end{document}